# Compressed Sensing Algorithms for OFDM Channel Estimation

Jonathan Ling, Dmitry Chizhik, A. Tulino, and Inaki Esnaola[2]
[1]Alcatel Lucent, Bell Laboratories, Wireless Communications Research
[2]University of Delaware, Electrical Engineering

*Abstract—* Radio channels are typically sparse in the delay domain, and ideal for compressed sensing. A new compressed sensing algorithm called eX-OMP is developed that yields performance similar to that of the optimal MMSE estimator. The new algorithm relies on a small amount additional data. Both eX-OMP and the MMSE estimator adaptively balance channel tracking and noise reduction. They perform better than simple estimators such as the linear-interpolator which fix this trade-off *a priori*. Some wideband measurements are examined, and the channels are found to be represented by a few delays.

## I. INTRODUCTION

Wireless transmission links use reference symbols or pilots so that transmitted data symbols can be properly demodulated. In the cellular standard LTE the overhead due to pilots is about 5% per antenna. Multiple transmit antennas can be used for diversity or improving received signal power or sending multiple spatial streams, all which increase transmission throughput. As the number of antennas grows so also does the overhead due to pilots. In the case of spatial multiplexing, overhead due to pilots actually limits the number of useful transmit antennas and consequently the throughput [1]. In the case of downlink beamforming at the basestation, all mobile users must make available their channels, usually with uplink pilots. Likewise, uplink pilot overhead can limit the number of orthogonal users [2].

Compressed sensing [3] for channel estimation relies on the assumption that channels can indeed be represented compactly in some basis, and thus fewer samples are required to learn the channel than what was traditionally thought. Traditional Nyquist sampling requires that the sampling rate be at least twice the highest frequency component of the signal (that we might be interested in), whereas compressed sensing (CS) theory tells us that the number of samples needed is ideally proportional to the amount of information in the signal. Note, while it is possible to estimate both data and pilots jointly, e.g. [4], this paper considers their estimation separately.

The sparseness of radio channels is the consequence of the environment and the necessary fidelity of the representation. Some channel models reduce the propagation environment down to a few significant scatterers, each generating a single plane wave having a certain azimuth and delay at the receiver. Such channels would appear to be highly compressible. In order not to predetermine the results we will assume that the channel is continuous in the delay domain. We have also assumed block-fading in time do not attempt to take advantage of potential compactness in the Doppler domain. One reason is that in current systems, packet transmissions are independent in time from another. In order to save power, the receiver partially shuts down until it knows there is data to receive, and so this behavior will limit its ability to track Doppler shifts.

Consider a single set of $d$ contiguous OFDM subcarriers, with $N$ pilots or reference symbols as a subset of these subcarriers. Now the pilots, here taken as unit gain symbols, are multiplied by the channel and received with thermal noise so

$$y = \mathbf{H}\theta + n, \qquad (1)$$

where $\mathbf{H}$ is $N$x$d$ DFT matrix, and $\theta$ is $d$x1 vector of model coefficients and $n$ is $N$x1 vector of additive white Gaussian measurement noise. The goal is to estimate $\theta$ given that $y$ has been observed. Given an estimate of $\theta$, it may be used to predict channels at the other frequencies (or subcarriers) that contain useful data rather than pilots. In the compressed sensing framework, the basis is clearly Fourier, and $\theta$ are the channel coefficients in the time-delay domain. It is often assumed that $\theta$ is $m$-sparse, i.e. $\theta$ contains no more than $m$ non-zero components. While there are no exactly sparse signals in nature, $\theta$ can be considered sparse after thresholding. How large $m$ actually is will be given for some measured channels.

One class of compressed sensing algorithms is called basis pursuit, which formulates the problem as a convex relation of the more difficult to solve l0 minimization. A particular form of basis pursuit is Lasso [5]:

$$\text{minimize} \|y - \mathbf{H}\theta\|_2$$
$$\text{subject to}: \|\theta\|_1 \leq c \qquad (2)$$

The l1 norm, which is the sum of the absolute values of $\theta$, is observed on the constraint, and this is what makes the solutions sparse. This estimator is Bayesian in the sense that $c$ is assumed known. Unfortunately there is no way to include any prior information one has on the individual elements of $\theta$, so our main focus will be on a class of greedy algorithms, described next.

A second class of algorithms is based on a greedy approach and called orthogonal matching pursuit (OMP). This algorithm determines the non-zero elements of $\theta$ by successively choosing the columns of $\mathbf{H}$ that are most likely to be in the model given the algorithm's prior selections. It was shown recently in [6] that the algorithm operating with random sensing matrices could in fact reconstruct the sparse signal with measurements in nearly linear proportion to signal sparsity *m*. Several variants of OMP have been proposed, e.g. [7][8]. A particular version called P-OMP is Bayesian in that it weights the selection of model components by probability that this component is part of the model [8]. Several researchers [17][18][19] have evaluated both basis pursuit and OMP, finding opportunities for reducing the number of pilots. In this paper, we attempt to improve the performance of OMP at low SINR, as these SINRs are most prevalent in cellular systems, by incorporating previously transmitted pilots into the estimation process.

The remainder of this paper is organized as follows. In Section II sparsity in the delay domain is analyzed using radio channel measurements. Section III discusses the MMSE estimator, which utilizes the perfectly known power delay profile. The performance of the MMSE estimator is used later as upper bound. Section IV exposes the difficulties using power-delay profile obtained by random sampling. Section V proposes some practical estimators that utilize priors. Section VI compares the performance of standard and CS estimators for OFDM with certain fixed number of pilots corresponding to LTE.

## II. SPARSITY IN TIME DELAY

For an OFDM communication system, it suffices to represent the channel as a delay line with *d* taps. Now if compressed sensing will reduce the pilot overhead or reduce estimation error depends on the channels. It is generally believed that they are sparse in the delay domain, i.e. there are just a few significant values of $\theta$ while the remainder of $\theta$ is nearly zero. Due to the fact that scatterers are relatively far from the receiver the 2nd order statistics and delays remain constant over a small area. The result is a process that is wide sense stationary, and the delays are conditionally independent. The channel is the sum of many plane waves, so $\theta$ has Gaussian statistics, and the covariance of $\theta$,

$$\mathbf{C}_\theta \equiv E\{\theta^T \theta\} \qquad (3)$$
$$S \equiv diag(C_\theta)$$

completely defines the process. The diagonal is defined as the power delay profile (PDP), and the expectation is over the spatial area where the model is deemed valid.

Two models that will be utilized in this paper are the extended-typical-urban model (ETU) [20] which specifies PDP with 9 impulses, and the 3GPP spatial-channel-model (SCM) with the far-scatter option [12] which specifies a PDP with 6 impulses. In order not to predetermine the outcome, the channel models are modified to fill all *d* delays, by converting the impulses into continuous PDP. Each impulse in the original model now becomes a set of exponentially decaying delays whose sum has the power of the original impulse. The RMS delay is set to 0.1 μS.

While the standard models are sufficiently representative for normal channel estimation algorithms, realizing the benefit of compressed sensing algorithms depends critically on how sparse channels are. Thus new analysis of measurements is necessary. Researchers often use rms delay spread, i.e. the variance of power weighted delays, as a single metric to characterize the frequency selectivity of the channel. However, high delay spread does not imply large *m*. On the other hand low delay spread does imply small *m*. The metric 95% *energy support* used in [16], has similar deficiency with respect to *m*. Nonetheless empirical evidence as illustrated in Figure 1, shows low measured delay spread, and most likely a small number of delays. Even so it is not obvious how many delays there are at higher bandwidth.

We report the number of delays $\eta_{95}$ necessary to capture 95% of the signal energy. The relative residue error is therefore -13dBc. Qualitatively, the channel estimation quality does not need to be that much better than the actual channel SINRs, which are mostly quite low. A typical simulation of geometries shows that 75% of the locations being less than 10 dB SINR with 3 degree downtilt. For higher SNRs more of the signal power needs to be captured leading to larger number of delays.

A small set of data from a peer-to-peer measurement campaign in a rural area taken at 2 GHz over a 6 MHz bandwidth [16] is analyzed. The quantity $\eta_{95}$ is plotted in Figure 2, versus the measured rms delay spread. There are two outliers, with very high delay spread and high $\eta_{95}$, which arise due to multiple reflections between aircraft hangers. In most cases however the number of delays is low (<15), on the order of what has been selected for 3GPP models, which at least partially validates the critical assumption we and others have been making. The number of delays also shows low correlation with delay spread. The line drawn is the number of delays is the profile where exponentially decaying. Channels to the right, with large delays but small $\eta_{95}$, appear to be better served by CS rather than algorithms based on uniform sampling.

The original ETU and SCM channel models specify a fixed number of delays irrespective of bandwidth or cell size. In cellular system, increasing bandwidth might tend to linearly increase the number of delays, but decreasing cell radii decreases the delay spread as the main arrival become more prominent, leading to a reducing in the delays. These are open issues which need to be investigated.

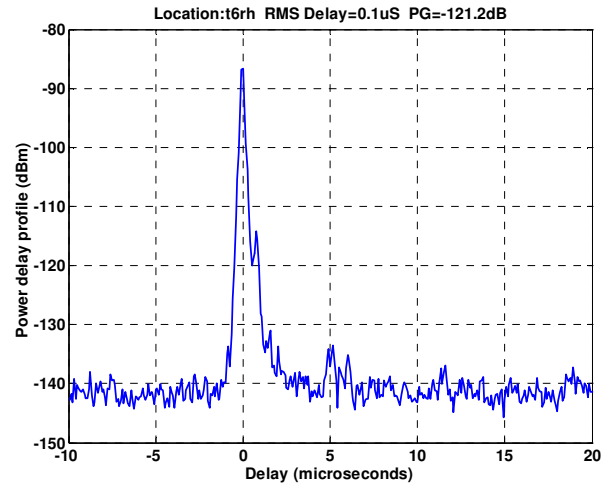

Figure 1. Example of a power delay profile with low delay spread and small number of taps.

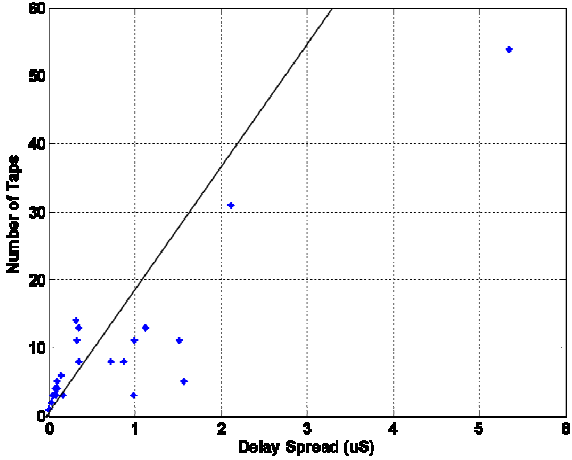

Figure 2. $\eta_{95}$ vs. measured delay spread with 6 MHz BW. The straight line, gives the number of taps to represent an exponential PDP with rms delay given by the x-coordinate.

## III. MMSE ESTIMATOR

The optimal linear Bayesian estimator when the PDP $S$ known exactly is discussed. Since the parameters are Gaussian, the estimator is also the MMSE estimator and achieves the Cramer-Rao lower bound (the lowest mean-square error of unbiased estimators). To develop a practical algorithm $S$ may need to be estimated, but having perfect knowledge serves as an upper bound on performance.

The estimator is the conditional expectation of $\theta$ given $y$, and for this discussion let $N$ equal $d$. A useful alternative form of the MMSE estimator for the linear model obtained by the matrix inversion formula is [11]

$$E(\theta | y) = \left(\mathbf{C}_\theta^{-1} + \mathbf{H}\mathbf{C}_{n_0}^{-1}\mathbf{H}^T\right)^{-1} \mathbf{H}^T \mathbf{C}_{n_0}^{-1} y = \mathbf{G}\mathbf{H}^T\mathbf{C}_{n_0}^{-1} y, \quad (4)$$

where $\sigma_{\theta,k}^2$ represents the $k$-th element diagonal covariance matrix $\mathbf{C}_\theta$. The parameter vector $\theta$ is length $d$ and not sparse. Assume that $\mathbf{H}\mathbf{C}_{n_0}^{-1}\mathbf{H}^T \approx N\mathbf{I}/\sigma_{n_0}^2$. On the $k$-th diagonal element of $\mathbf{G}$ the is

$$\mathbf{G}_k = \left(\frac{1}{\sigma_{\theta,k}^2} + \frac{N}{\sigma_{n_0}^2}\right)^{-1} \quad (5)$$

Consider the following two cases: where the PDP on the given delay is much greater than, or much less than the noise.

$$\begin{aligned}\sigma_{\theta,k}^2 \gg \sigma_{n_0}^2 &\to \mathbf{G}_k \approx \sigma_{\theta,k}^2 \\ \sigma_{\theta,k}^2 \ll \sigma_{n_0}^2 &\to \mathbf{G}_k \approx N/\sigma_{n_0}^2\end{aligned} \quad (6)$$

While the MMSE is always full rank, we can consider a reduced rank model of significant channel taps. Let M contain the set of $m$ column indices, and $\mathbf{H}_M$ the eigenvectors of this model. The reduced-rank estimator becomes

$$\theta_M = \mathbf{G}_M \mathbf{H}_M^T y / N \approx \mathbf{H}_M^T y / N, \quad (7)$$

where the last approximation is useful at moderate and higher effective SNRs. The noise in the resulting estimate is reduced by a factor of

$$K = \frac{N}{m} \quad (8)$$

compared to an $N$ dimensionality estimator, such as least-squares. In summary the MMSE has knowledge that is generally unavailable, and utilizes it optimally.

The estimator (7) suggests a two-step process whereby the support, i.e. the set of significant channel taps, are determined followed by least-squares estimation. In fact this will be the basis of the algorithms of Section V. Indeed such a procedure attempts to obtain an optimal representation. For a discretely sampled process the eigenvectors of its covariance matrix are the orthonormal functions of the Karhunen-Loeve expansion. The covariance matrix of our random channel is

$$\begin{aligned}R(f,f) &= F\{R(\tau,\tau)\} \\ &= \sqrt[-1]{N}\mathbf{H}\mathbf{C}_\theta \sqrt[-1]{N}\mathbf{H}^T = \sum_{n=1}^N \theta_n \mathbf{H}_n \mathbf{H}_n^T / N\end{aligned} \quad (9)$$

Thus conditioned on a specific realization of channel taps, i.e. fixing their location and variance, the DFT is the Karhunen-Loeve expansion. In the noiseless case, since the eigenvalues are i.i.d., we can represent the process to any fidelity, as the variance of the sum is the sum of the variances, starting with the largest eigenvalue. The noise is assumed white and therefore appears with equal power in all dimensions. Thus the model consists of all eigenvectors whose associated eigenvalues are a certain threshold relative to noise. Clearly the output SNR ceases to improve when the next largest eigenvalue has less signal power than the noise. OMP and its variants attempt to select the correct set of eigenvectors in the presence of noise. Utilizing the sample PDP obtained from multiple sets of data, correlated or uncorrelated, improves the probability of making correct decisions.

## IV. OBTAINING THE PRIORS VIA RANDOM SAMPLING

As a prequel to proposing practical algorithms, we consider the task of estimating the PDP and distinguishing between signal and noise for randomly placed pilots. The compressed sensing framework requires a special sort of sampling, such that the columns of the sensing matrix $\mathbf{H}$ are nearly orthogonal [3]. The Fourier matrix with randomly chosen columns meets this criterion. To obtain priors, i.e. a PDP estimate, one must use the available random samples. It has been shown that in contrast to uniform sampling where aliasing occurs if the signal frequency is above ½ the sampling rate, random sampling has no such limitation [9][10].

The disadvantage of random sampling is that strong leakage shows up in all frequency bins. In fact even in the noiseless case the average power of the leakage is equal to the signal power of itself. Here, one can think of the inner product of a random vector and the measurement vector. A single tone will have a coherent gain of $N^2$ at its true frequency, and gain of $N$ at all others. For more general signals, weak components will be masked by the leakage, see also [14, pg. 383]. For example, Table 1 gives the ETU PDP, which shows that the later delays are weaker than the rest, and would be therefore more difficult to detect.

There are two ways to improve the PDP estimate, the method depending on whether the data are correlated or not. Assume that there are $N_a$ sets of $N$ pseudo-randomly distributed pilots. When these sets are completely correlated, i.e. they experience exactly the same propagation channel, the can be processed together effectively increasing $N$ by factor of $N_a$. However pilots are often uncorrelated or nearly uncorrelated due to time variations in the channel, thus incoherent averaging must be performed.

Consider the sample power delay profile at $\tau$.

$$\hat{S}(\tau) = \frac{1}{N_a} \sum_{n}^{N_a} |\theta_S^{(n)}(\tau)|^2 \qquad (10)$$

Define the random variable

$$S_I = \hat{S}(\tau), \quad \tau \in I \qquad (11)$$

where $I$ is the set of delays that do not contain signal. Since $x_s(\tau)$ is complex Gaussian by the central-limit theorem; therefore $S_I$ is a Chi-square random variable with $2N_a$ degrees of freedom and

$$\begin{aligned}\mu_{S_I} &= \left(\sigma_x^2 + \sigma_{n_0}^2\right)/N \\ \sigma_{S_I}^2 &= \left(\sigma_x^2 + \sigma_{n_0}^2\right)/N \cdot 2N_a\end{aligned} \qquad (12)$$

Thus both incoherent and coherent averaging will improve peak detection.

Without any knowledge of the distribution of the signal power itself, using hypothesis testing, one chooses a maximum allowable false alarm rate, i.e. largest number of false positives we are able to accept [14]. Therefore a peak is declared a signal component (part of the signal support), if

$$\hat{S}(\tau) > F_{S_I}^{-1}(1-\alpha) \qquad (13)$$

where α is the false alarm rate.

Figure 3 shows $\hat{S}$ with $N=80$, and $N_a=5$, at 0 dB SNR. Based on the figure, it is difficult to make a determination between signal peaks and spurious noise. If a low false alarm rate is set some signal peaks will be missed. In particular the signal component at 5 μs is not discernable, and will apparently always be missed.

Table 1: ETU power delay profile from [20].

| Delay (nS) | Relative Power (dB) |
|---|---|
| 0 | -1 |
| 50 | -1 |
| 120 | -1 |
| 200 | 0 |
| 230 | 0 |
| 500 | 0 |
| 1600 | -3 |
| 2300 | -5 |
| 5000 | -7 |

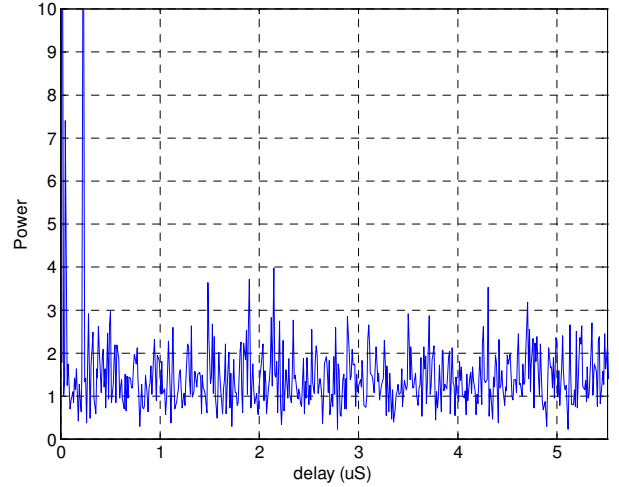

Figure 3: $\hat{S}$ with $N=80$, and $N_a=5$, 0 dB SNR.

V. PRACTICAL ESTIMATORS USING THE PRIOR

The reader is directed to [6] for a full description of orthogonal matching pursuit (OMP), which is the basis of algorithms A2, A3, and eX-OMP. Each of the new algorithms utilize $N_a$ uncorrelated (or mostly uncorrelated) sets of $N$ pseudo-randomly pilots.

A1: Form the sample PDP (10) given $N_a$ sets of data. Set α=0.001, and determine set M of col. indices by (13), and perform (7).

In the following algorithms the column selection operation of OMP is modified. Specifically in the $i$-th iteration OMP determines a new column $m^{(i)}$, using the residue $r^{(i)}$ which holds the remaining un-modeled signal and noise, by projecting it onto the basis

$$m^{(i)} = \arg\max\left\{\mathbf{GH}^T r^{(i)}\right\}, \qquad (14)$$

where $\mathbf{G}$ is the identity matrix unless otherwise specified.

A2: Introduce MMSE-like weighting into the col. selection process of OMP. Let $\mathbf{\Lambda}_p$ represent the thresholded PDP. By setting a threshold we attempt to improve the SINR of the PDP (10). If a power at a certain delay is below the threshold it is said to be noise, and assigned the average noise value. Now for the $i$-th iteration of OMP let $\mathbf{\Lambda}_r^{(i)}$ represent the variance of the noise and the leakage in the residue r. The weighting matrix is

$$\mathbf{G}^{(i)} = \left(\mathbf{\Lambda}_P^{-1} + \mathbf{\Lambda}_r^{(i)\,-1}\right)^{-1} \qquad (15)$$

Note that $\mathbf{\Lambda}_p$ remains fixed while $\mathbf{\Lambda}_r^{(i)}$ decrease each iteration.

A3: Form the sample PDP (10) given $N_a$ sets of data. Set α=0.001, and determine signal peaks by (13). The peaks are taken as $m^{(0)}$, rather than (14). OMP iterations continue regularly thereafter.

eX-OMP: Consider $N_a$ OMPs each operating independently on their own set of $N$ pseudo-random pilots. Instead of (14) perform the following:

1. Exchange and combine residues $r$ at each iteration to form a single PDP via incoherent addition.
2. Threshold by (13) to admit potentially multiple columns.
   If none are selected, then chose the column corresponding to the largest delay.

Thus a single set of delays for all OMPs is created. Estimation and subtraction are performed as usual individually. Note that step 2, is similar to StOMP [7]. The complexity of this algorithm is O($N_a mNd$). Least squares can be solved in O($tN$) where $t$ is the iteration. eX-OMP can also be used to efficiently solve the channel estimation of a MIMO receiver.

## IV. APPLICATION TO LTE

The LTE downlink is characterized by regularly spaced pilots, spaced every third subcarrier. Three standard estimation [**Error! Reference source not found.**, pg. 170] algorithms: DFT, linear interpolator (LI), and linear interpolator with MMSE (LI-MMSE) will be compared to the new algorithms of Section V, and the MMSE estimator. Introducing some notation, let P be the set of indices of the pilot symbols with cardinality $N$. Let D be the set of indices associated with the data subcarriers.

Given the linear model (1), the DFT approach first estimates the delays via the IDFT

$$\bar{\theta}_P = \mathbf{H}_P^T y_P, \qquad (16)$$

where $\hat{\theta}_P$ is $N$x1 vector and $\mathbf{H}_P$ contains the rows of $\mathbf{H}$ which correspond to the pilot subcarriers. The channel estimates at the data symbols are found via the DFT

$$\hat{y}_D = \mathbf{H}_D \bar{\theta}_P = \mathbf{H}_D \mathbf{H}_P^T y_P, \qquad (17)$$

where $\mathbf{H}_D$ contains the rows of $\mathbf{H}$ which correspond to the desired data subcarriers. Note that since $N<d$, and with P uniformly spaced, we have upper limit on detectable delays, before aliasing occurs. In LTE with 15 KHz subcarriers, and for $d$=600 and $N$=200, aliasing occurs when delays are greater than 11μS. This estimator provides no noise reduction; however increasing $N$ increases the largest resolvable delay, which might otherwise contribute to the effective noise-floor.

The linear interpolator operates in the frequency domain. The channel of data symbols is simply a linear estimate of the two adjoining pilot symbols. This estimator is very easy to implement, and assumes that the channel is smoothly varying between the pilots. The best noise gain is a factor of 2 or 3dB, as each interpolated frequency is the linear weight of two pilots.

Another straightforward estimator is LI-MMSE, which takes the noise variance (known perfectly), and the sample covariance matrix formed from previously transmitted pilots. This estimator is

$$\hat{y}_P = \hat{\mathbf{C}}_P \left( \hat{\mathbf{C}}_P + \sigma_{n_0}^2 \mathbf{I} \right)^{-1} y_P \qquad (18)$$

where $\hat{\mathbf{C}}_P$ is low resolution with dimensionality $N$x$N$ rather than $d$x$d$. Thereafter the LI is used to find the channel at the data symbols. Estimation of $\hat{\mathbf{C}}_P$ is full rank, and therefore without any noise reduction. Incoherent averaging improves the statistical estimate, but not the SNR. Therefore, no noise reduction is possible beyond that of the LI. Clearly, if the true covariance matrix were available, there would be noise reduction.

To evaluate all the algorithms, the normalized mean square error (NMSE),

$$10 \log_{10} \frac{E\left\{ |y_D - \hat{y}_D|^2 \right\}}{\|\theta\|^2} \text{ (dB)} \qquad (19)$$

where the expectation is obtained by averaging over many realizations of the channel, is used as a metric. The performance of the standard algorithms is given in Figure 4. The NMSE of DFT approach is the same as the SNR itself. The linear interpolator actually does 2.5 dB better than the DFT at low SNR, but does exhibit an error floor observed at higher SNRs due the fact that the channel isn't really varying linearly between pilots. Using MMSE in combination with LI, improves the performance at low SNR as expected. The MMSE performs about 9 dB better than DFT due to the smaller number of parameters in the model, i.e. (8).

The goal of the Bayesian compressed sensing algorithms, is to be as close as possible to the MMSE estimator's performance. The amount of prior data is set at $N_c$=8. Figure 5 given the performance of the Bayesian estimators along with OMP. OMP in this case provides about 6 dB of noise reduction. Algorithm A1 estimates the parameters found by hypothesis testing. At higher SNRs the NMSE saturates at -17 dB due to its inability to detect weak peaks. So while A1 is simple, the tradeoff is that it is limited to low operating SNRs. The new OMP based algorithms A2 and A3 provide some improvement, but eX-OMP's is even better. Both A2 and A3 take advantage of the readily discernable peaks to give OMP a head start, which produces the desired improvement over OMP at low SNRs. In A3 $\Lambda_r^{(i)}$ becomes smaller with each iteration, so the effect of the prior eventually becomes insignificant. eX-OMP is at most less than 1 dB from the MMSE and makes better use of the uncorrelated priors. Only eX-OMP continues to use all the data by subtracting the known signal to reveal the weaker peaks. Lasso, not shown, performs on par with OMP. The true value of $c$ is used on a realization basis as suggested in [13]. Analysis shows that Lasso tends to produce models that are not sparse enough. It is possible to improve Lasso by optimizing over $c$, if the true channel is also known, but if there is an optimal way to set it *a priori* is an open question.

The results thus far have been expressed in terms of NMSE, and to understand how it affects rate, we use the model of Hassibi and Hochwald. The model requires that the channel estimate be uncorrelated with the error. Since OMP employs least squares, the error is uncorrelated with the estimate. A lower bound on the ergodic capacity accounting for the estimate error is [1, eq. 20]

$$C_l = \frac{(N_s - N)}{N_s} E\left\{ \log_2 \left( 1 + \rho \frac{1 - \sigma_e^2}{1 + \rho \sigma_e^2} \theta^T \theta \right) \right\} \qquad (20)$$

where $\rho$ is the SNR, $\sigma_e^2$ is the NMSE, $N_s$ is the number of symbols in a coherence interval, and the expectation is taken over the realizations of $\theta$. Pilot spacing is adjusted to be every 1/6 subcarrier, for 8×8 MIMO. Figure 6 shows the fraction of capacity achieved vs. SNR for the ideal, linear-interpolator, and eX-OMP channel estimators. The linear-interpolator fares much worse at $N$=100 than for $N$=200, making the DFT a better choice. Nonetheless eX-OMP substantially improves capacity at the lower SNRs.

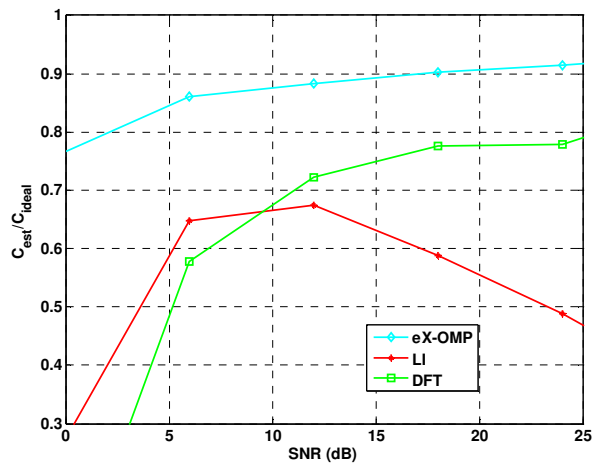

Figure 6: Fraction of capacity vs. SNR for ideal and practical channel estimators with $N$=100.

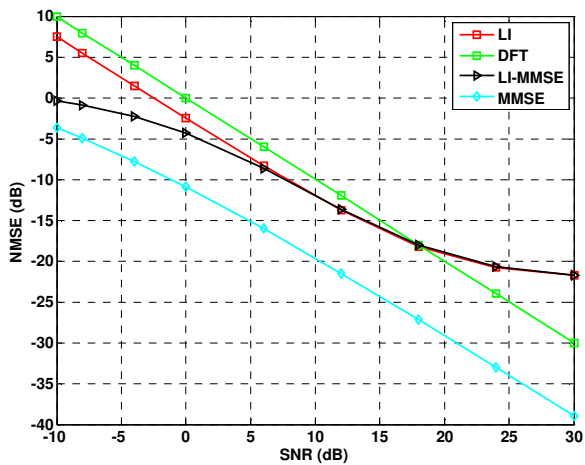

Figure 4: Performance of standard channel estimation algorithms with $N$=200 for the SCM channel model.

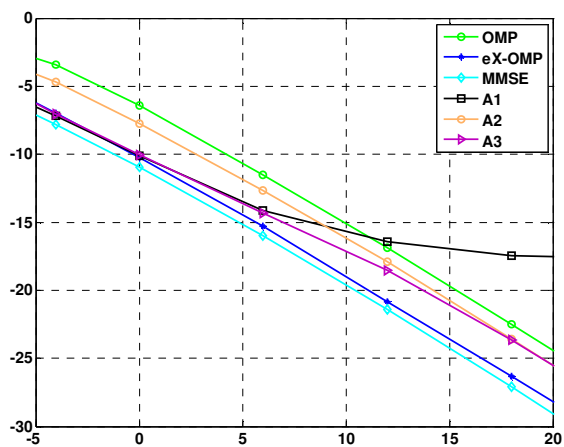

Figure 5: Performance of Bayesian estimators and OMP. $N$=200, $N_a$=8, for the ETU channel model.

## IV CONCLUSIONS

A limited number of wideband measurements were examined, and found to be represented by a small number of delays, even for cases of higher delay spread. These channels are therefore ideal for the compressed sensing approach. A new Bayesian compressed sensing algorithm called eX-OMP was developed. The new algorithm relies on additional data. For the range of SNR applicable to cellular systems, numerical experiments show that only 8 additional sets of pilots are sufficient to yield performance of the MMSE estimator, which optimally trades off channel-tracking and noise-reduction. eX-OMP performs better than simple estimators such as the linear-interpolator or moving-average that incorporate a fixed trade-off between noise and channel tracking capability.

## ACKNOWLEDGEMENTS

We would like to thank Reinaldo Valenzuela and Sivarama Venkatesan for discussions on channel modeling and performance evaluation.